\documentclass[12pt]{article}
\usepackage{graphicx,subfigure,psfrag,url} 
\usepackage{url,cite,setspace}

\makeatletter \renewcommand\@biblabel[1]{} \makeatother

\begin{document} 
\title{Dynamic Moment Analysis of the Extracellular Electric Field of
a Biologically Realistic Spiking Neuron} 
\author{
Joshua N. Milstein\\
\texttt{milstein@caltech.edu}\\
California Institute of Technology\\
Pasadena, CA 91125\\
\\
Christof Koch\\
\texttt{koch@klab.caltech.edu}\\
California Institute of Technology\\
Pasadena, CA 91125\\
}
\maketitle

\begin{abstract}
\noindent Based upon the membrane currents generated by an action potential in 
a biologically realistic model of a pyramidal, hippocampal cell within
rat CA1, we perform a moment expansion of the extracellular field
potential.  We decompose the
potential into both inverse and classical moments and show that this
method is a rapid and efficient way to calculate the extracellular
field both near and far from the cell body.  The action potential
gives rise to a large quadrupole moment that contributes to the
extracellular field up to distances of almost 1 cm.  This method will
serve as a starting point in connecting the microscopic generation of
electric fields at the level of neurons to macroscopic observables
such as the local field potential.

\end{abstract}

\section{Introduction}

Since the pioneering work of Hodgkin and Huxley in the early fifties
\cite{hodgkin} on the initiation and propagation of action potentials
within the squid giant axon, there has been significant progress in
our understanding of brain function at the level of the single neuron
\cite{koch}.  Unfortunately, it has proved difficult to connect
function at this microscopic scale to more global, large-scale brain
function.  In this paper, we work toward this goal by developing a
physiologically accurate model of the extracellular field of a single
neuron which may be efficiently employed to model the field associated
with very large numbers of neurons.

The dominant means of rapid communication among neurons is through
chemically or electrically mediated synapses.  Ephatic interactions, where
communication is directly via an electric field, may occur in nerves
that have been crushed or damaged by neurodegenerative disorders such
as multiple sclerosis \cite{faber1,jefferys}, but examples of ephatic
effects under normal conditions are rare \cite{kandf,kanda}.
Nonetheless, all electronic, cellular activity generates extracellular
electric fields and so it is natural to ask if these fields have any
relevance to the functioning of the brain.  Before we can begin to
answer this question, however, we need to consider how best to model
these fields.

Our current objective is to better understand the forward problem of
modeling the extracellular field of various regions of the brain from
the underlying, neural activity and to develop an accurate and
efficient method for modeling these fields.  A full construction of
the extracellular field, from single neuron activity, is extremely
difficult.  For instance, to generate microvolt potentials, as
commonly detected by electroencephalograph (EEG) scalp recordings,
requires the superposition of activity from a great number of neurons.
A simple estimate is that it takes a $6\ {\rm cm}^2$ patch of
cortical tissue, containing around $6 \times 10^{7}$ synchronously
active neurons, to generate a detectable signal on the order of
microvolts \cite{ebersole}.  Nonetheless, the microscopic behavior,
although too difficult to incorporate exactly, may act as a guide in
developing more coarse-grained models \cite{srinivasan}.  For
instance, field theories of thalamic and cortical activity,
constrained by physiological parameters, have recently been developed,
and have proven successful in quantitatively reproducing various EEG
phenomena, evoked response potentials, coherence functions and seizure
dynamics, among others.  \cite{jirsa,rob1,rob2}.

Neurons display a variety of complicated geometries, giving rise to an
array of current distributions that dynamically vary throughout the
course of an action potential and during the interspike interval.  For
local probes of individual neurons--for instance, by
microelectrodes--the field generated by the action potential
dominates, particularly near the soma.  However, it is thought that
synaptic activity as well as longer lasting depolarization and
hyperpolarizations are mainly responsible for the electrical activity
detected by EEG recordings \cite{nunez}.  There are two primary
reasons why the contribution of the action potential is thought to be
negligible to the fields detected by EEGs. First, in general, the
dendritic axes of pyramidal cells lie parallel to the cortical sheet
which allows the contribution of the extracellular fields of the
dendrites to add constructively, whereas the relative orientation of
their axons are more varied, leading to a significantly reduced axonal
contribution. Second, due to the relatively brief time course of an
action potential, neurons would need to precisely synchronize their
firing in order to generate a significant contribution to the
extracellular field.

In the current study, we focus on the extracellular field of a single
spiking cell, with the future intention of quantifying hypotheses such
as those discussed above on the importance of the action potential to the
extracellular field. We base our
work upon a quantitatively accurate model of a pyramidal cell which
our lab has developed and  use this model to ask
questions regarding the local extracellular field, for instance, the
field generated by a single neuron or a minicolumn of pyramidal
neurons, and later address how our results are relevant to more
distant, global recordings, such as EEGs.

The dynamics of the extracellular field of a spiking neuron are rather
complex.  One would like to remove some of the complexity of analyzing
the extracellular field of realistic neurons by identifying the
essential features that characterize the current distributions.  With
this intention, our approach is to perform a moment expansion about
the current distribution of the cell and to study the resulting,
dynamical moments.  Moment expansions are routinely used in molecular
biology to aid in the calculation of Coulomb mediated molecular
interactions where the full electrostatic charge density may be quite
complicated. They have been used to clarify the possible interactions
between normal and alkylated DNA base pairs \cite{price}, to model
ligand binding and protein-protein interactions \cite{petersen}, and
to simulate charge transport in biological ion channels
\cite{saraniti}, to name only a few applications.  Our present goal is
two-fold: to first show that the dynamical moments of a biologically
realistic neuron can be efficiently calculated and to then see what
simplifying features emerge from such an analysis.

Our current approach naturally leads to several fundamental questions
which have not been sufficiently addressed: when is
it justified to model the neuron by a dipole; is there a region of
interest where the first few moments provide a useful approximation to
the extracellular field; close to the cell, do any of the moments
dominate, or must we account for the full complexity of the current
distribution?  We present a method that is able to accurately and efficiently
decompose the extracellular field into its fundamental moments at all
distances from the cell body.  We then discuss the usefulness of such
an approach in describing local and global extracellular fields
generated by networks of neurons.

\section{Generalized Multipole Expansion}

We begin by writing an equation for the extracellular field of a
continuous source of currents within the point-source approximation
\begin{equation}
\phi({\bf x})=\frac{1}{4\pi\sigma}\int d^3 x' \frac{i({\bf x}')}{|{\bf
x}-{\bf x}'|},
\end{equation}
where $i({\bf x}')$ is the current at location ${\bf x'}$ and ${\bf
x}-{\bf x}'$ defines a vector which points from the current source
toward a test point at ${\bf x}$.  We will assume that the extracellular medium may be approximated as an
homogeneous, isotropic volume conductor and, therefore, the bulk
conductivity tensor $\sigma$, may be taken as a constant.  For
frequency ranges between roughly $1-3000$ Hz, capacitive effects are
negligible and a purely ohmic conductivity is sufficient for modeling
the extracellular milieu.  The validity of this approximation is
discussed in detail within references
\cite{holt,plonsey,destexthe}. Typical values of the bulk conductivity
range between $200-400\ \Omega\cdot{\rm cm}$.
  
Since we are interested in a multipole expansion of the
cell's current distribution at all distances from the cell, we need to
pay particular attention to the convergence properties of our
expansion method.  The usual decomposition into multipoles is based upon the
following expansion of the $1/|{\bf x}-{\bf x}'|$ dependence of the
electric potential into radial components $r$ and spherical harmonics
$Y_{l,m}(\theta,\phi)$:
\begin{equation}
\frac{1}{|{\bf x}-{\bf x}'|} =
\sum_{l=0}^{\infty}\sum_{m=-l}^{l}\frac{1}{2l+1}\frac{r_{<}^{l}}{r_{>}^{l+1}}Y_{l,m}^*(\theta',\varphi')Y_{l,m}(\theta,\varphi).
\end{equation}
The symbol $r_<$ refers to the smaller of the two values of $|{\bf
x}|$ and $|{\bf x'}|$ (for instance, ${\bf x}$ may be the vector which
points to the test point while ${\bf x'}$ points to the current
source), while $r_>$ refers to the greater value.  This condition will
insure that the sum is convergent, so special care needs to be taken
to abide by this criterion.  The classical multipole expansion assumes
that we are outside the range of the current distribution, so we may
identify $r_<$ with the magnitude of the vector pointing at current source
$r'$, while $r_>$ is associated with a test point at $r$.  However,
due to the complicated geometry displayed by different neurons, we may
easily find ourselves within a regime in which the identities of these
two quantities are swapped.

\begin{figure}[t]
\begin{center}
\includegraphics[width=0.5\columnwidth]{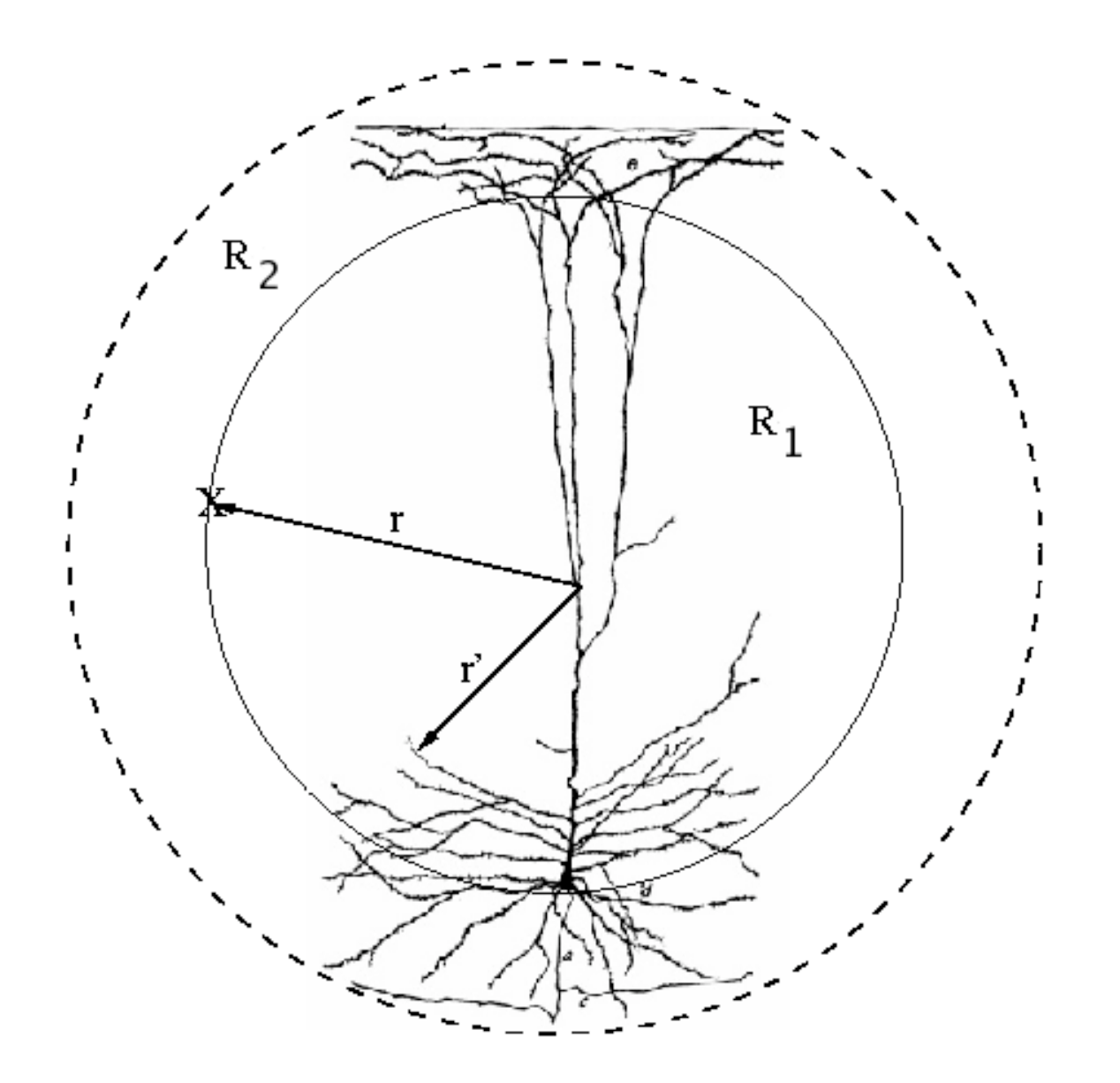}
\caption{The figure depicts the various regions into which the
generalized multipole expansion is divided.  Vector {\bf r'} points to
the current source while the vector {\bf r} is directed toward the
test point.  The solid circle divides
region $R_2$ from $R_1$.  The dashed line marks the divide between
the inner (inside) and outer (outside) field regions.  The pyramidal cell
depicted is purely illustrative.  }
\label{diagram1}
\end{center}
\end{figure}

Figure (\ref{diagram1}) clarifies this point.  We divide the
extracellular region of a stereotypical pyramidal cell into 2 distinct
volumes.  For convenience, we pick a point roughly halfway up the
apical dendrite of the cell as our origin; however, this choice is
arbitrary--for instance, we could have chosen the origin to fall
within the soma.  Our choice of origin simply minimizes the total
spherical volume of the current containing region, which will later
aid in the numerics.  For a test point at ${\bf r}$, the region $R_1$
denotes the volume over which $r_<=r',r_>=r$ while region $R_2$ is the
volume where $r_<=r,r_>=r'$.  The solid line separates these two
regions.  It's clear that for any value of ${\bf r}$ where there still
exists an element of current outside the volume enclosed by that
vector, we need to be careful that we have properly identified $r_<$
and $r_>$.  This leads to a natural splitting of extracellular space
into two regions, which is denoted by the dashed line in the figure.
We will refer to a test point within the volume enclosed by the dashed
line as in the ``inner-field,'' while points outside will be
considered the ``outer-field.''  We employ this terminology since the
regions we are considering are somewhat different than the more
typically encountered ``near'' and ``far'' field.  The important
distinction between this definition of an inner and outer-field is
that the outer-field defines the region in which $\{r_<,r_>\}$ are
static whereas, within the inner-field, $\{r_<,r_>\}$ vary based on
the placement of the test point.  For instance, for scalp recordings
several centimeters from the relevant cells, one is within the
outer-field, but for intracranial recordings millimeters from a
cortical microcolumn, one might have to account for the inner-field
based on the position of the electrode.

We may now write the following moment expansion of the extracellular
potential:
\begin{equation}\label{mexp}
\phi({\bf
x})=\frac{1}{\sigma}\sum_{l=0}^{\infty}\sum_{m=-l}^{l}Y_{l,m}(\theta,\varphi)\left(\frac{q_{l,m}}{r^{l+1}}+r^l
p_{l,m} \right),
\end{equation}
where
\begin{eqnarray}
q_{l,m} &=& \int_{R_1} d^3x'i({ \bf
x}')r'^lY_{l,m}^*(\theta',\varphi')\label{moments1}\\
p_{l,m} &=&
\int_{R_2} d^3x'\frac{i({ \bf
x}')}{r'^{l+1}}Y_{l,m}^*(\theta',\varphi')\label{moments2}.
\end{eqnarray}
Equations (\ref{moments1},\ref{moments2}) are the moments of the
potential, $q_{l,m} $ are the classical multipole moments, while
$p_{l,m}$ are the less familiar inverse moments.  If we write the
elements of the multipole expansion as \mbox{$\phi_{l,m}({\bf
x})=Y_{l,m}(\theta,\varphi)(\phi_{q_{l,m}}(r)+\phi_{p_{l,m}}(r))$},
from Eq.~(\ref{mexp}) we may define the classical and inverse radial
potentials
\begin{eqnarray}
\phi_{q_{l,m}}(r)\equiv \frac{1}{\sigma}\frac{q_{l,m}}{r^{l+1}}&{\rm and}&
\phi_{p_{l,m}}(r)\equiv \frac{1}{\sigma}p_{l,m} r^l,
\end{eqnarray}
respectively.  The radial potentials will be helpful in comparing the
relative importance of the moments in the multipole expansion. Note the radial
dependencies in Eq.~(\ref{mexp}) that guarantee convergence of the
expansion.  In the outer-field, this simply reduces to the standard
multipole expansion, but the series remains convergent within the
inner-field as well, so long as we restrict our integration over the
appropriate volume elements as denoted in
Eqs.~(\ref{moments1},\ref{moments2}) and illustrated in
Fig.~(\ref{diagram1}).  A
similar approach has recently been used to study the electrostatic
potential of topological atoms, from which we have borrowed some of
our terminology \cite{rafat}.

\begin{figure}[t]
\begin{center}
\includegraphics[width=0.75\columnwidth]{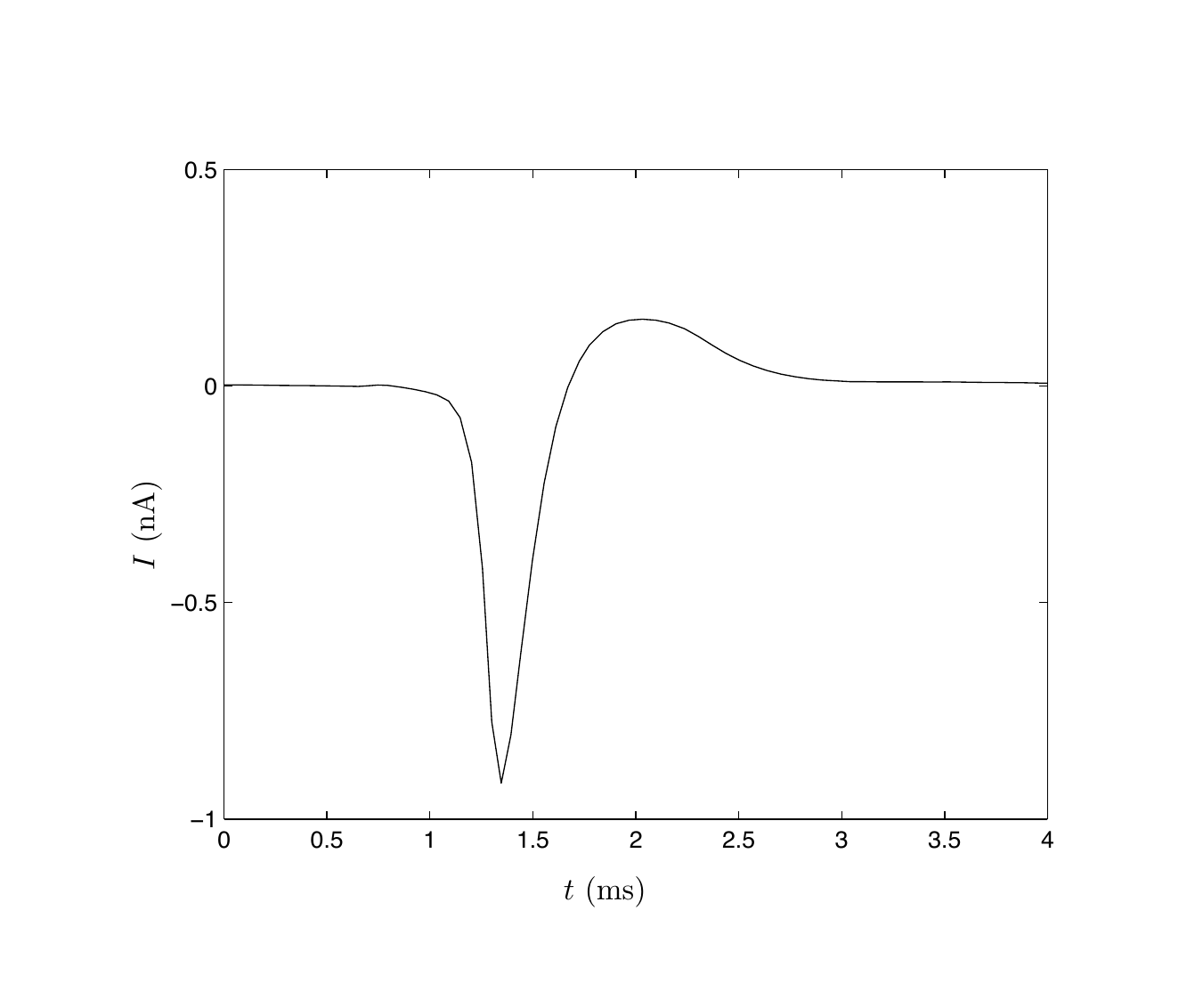}
\caption{A representative time course of the total current across the
soma showing the rapid inward (negative) ${\rm Na^+}$ current, leading to the
peak in the action potential, and the slower, outward (positive) ${\rm K^+}$
current which repolarizes the cell.  Simulated synaptic input occurs
within the first $1\ {\rm ms}$ triggering the firing of an action
potential.}
\label{somacurrf}
\end{center}
\end{figure}

\section{Model cell}

The cell that we will work with is a biologically realistic model of a
hippocampal pyramidal cell within rat CA1.  The model was developed in
 \cite{gold} to compare intracellular recordings to
simultaneous extracellular recordings of neural activity.  The active
ionic currents were modeled using Hodgkin-Huxley style kinetics.
Voltage dependent currents were carried by ${\rm Na}^+$, ${\rm K}^+$,
and ${\rm Ca}^{2+}$ ions and were modeled for 12 different current
processes.  Details of the model can be found in
\cite{gold}.  To calculate the extracellular field, we first computed
the transmembrane currents for the neuron along with their associated
ionic currents. Standard 1-D compartmental simulations where
performed within the NEURON Simulation Environment \cite{handc}.
Approximately 1000 compartments where used to model an anatomically
correct 3-D reconstruction of the cell.

\begin{figure}[t]
\begin{center}
\includegraphics[width=0.75\columnwidth]{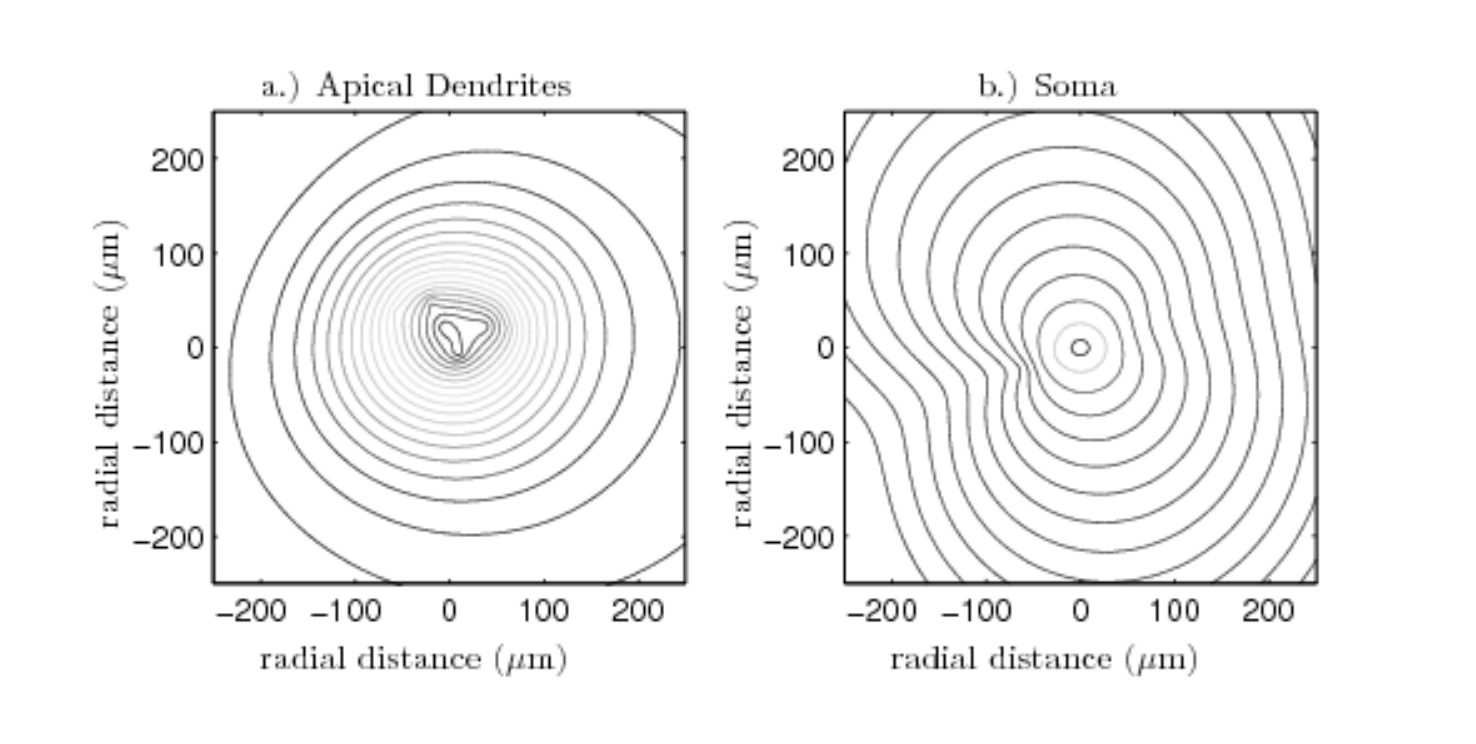}
\caption{Equipotential curves taken at the peak of the action
potential, calculated from the original pyramidal cell, illustrating
the approximate cylindrical symmetry of the extracellular potential.
We plot two cases above: a.) a plane at $z=250\ \mu m$, within the
apical dendrites and b.) a plane at $z=0$ which is the location of the
soma.}
\label{cylf}
\end{center}
\end{figure}

Within the first $ 1 \rm{ms}$ of the simulation we artifically
depolarize the cell until an action potential is triggered within the
soma; the cell dynamics follow the course of the action potential
until the cell repolarizes and returns to a stable resting potential.
This choice of initiating the action potential is arbitrary, we
could likewise apply the procedure discussed here to a cell whose
firing is initiated by synaptic input.  Figure (\ref{somacurrf}) shows
the time course of the membrane current across a representative
segment of the soma.  Throughout, we assume that the extracellular
potential is constant and equal to zero.  We also assume that the
transmembrane currents are not influenced by the evolving
extracellular potentials ($\ll 1 {\rm mV}$).  An iterative procedure
could be used to improve upon this approximation, although the
modification can be shown to be negligible \cite{holt}.

To study the moments that generate the extracellular field of a
pyramidal cell, we take advantage of the fact that theses cells are
almost, but not quite, cylindrically symmetric (see Fig.~(\ref{cylf})).
As a first pass, we assume that any anisotropy coming from the
branched structure of the dendrites is unimportant.  Assuming
cylindrical symmetry allows us to reduce the dimensionality of the
system from a three-dimensional calculation to a problem of only
two-dimensions, but should only modify the quantitative, as opposed to
qualitative, aspects of our results.

\begin{figure}[t]
\begin{center}
\includegraphics[width=0.8\columnwidth]{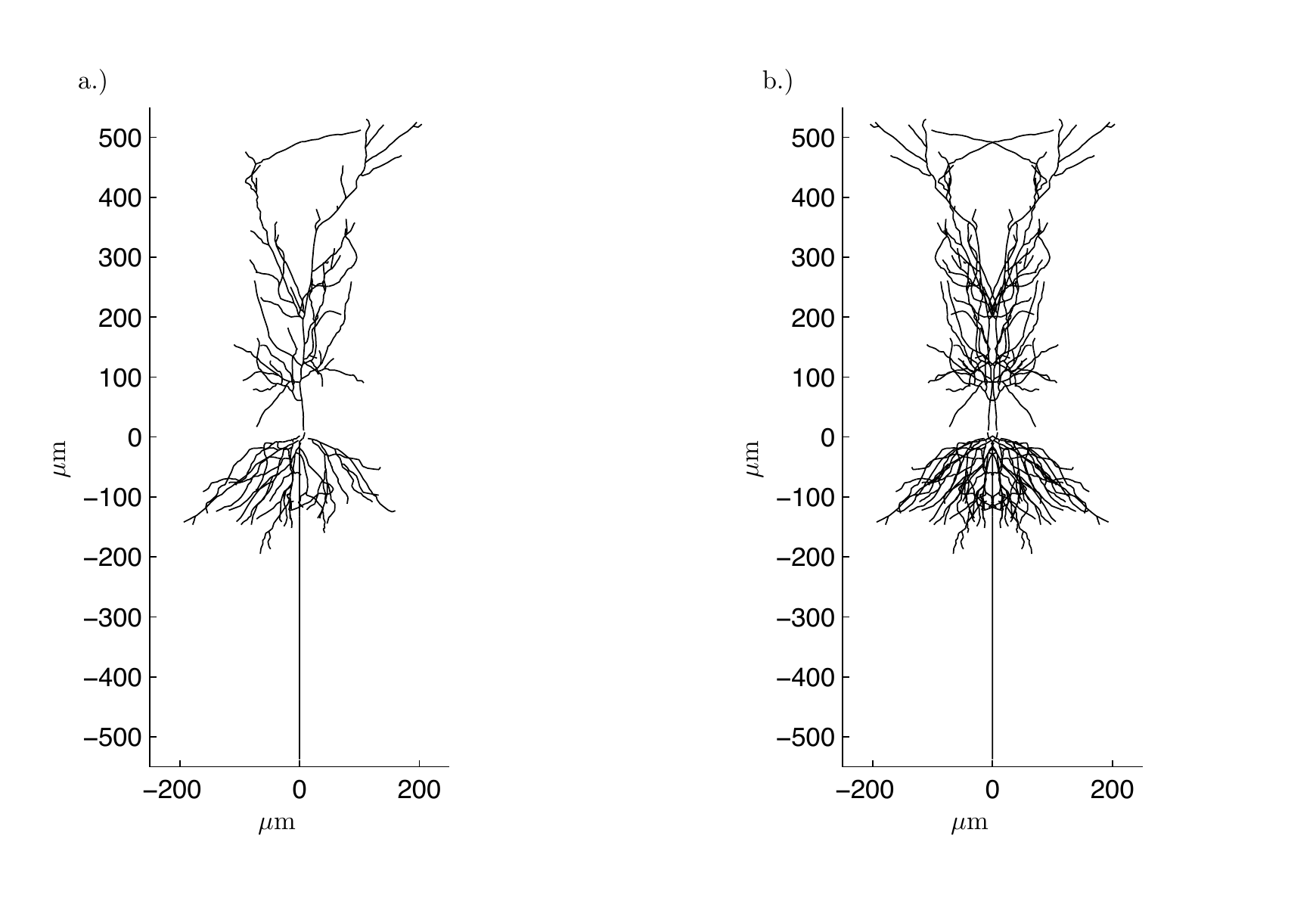} 
\caption{a.) A projection of the pyramidal neuron onto a plane
      perpendicular to the cortical section (left panel).  b.) The
      original cell is symmetrized to simplify the analysis (right
      panel).}
\label{celllayout}
\end{center}
\end{figure}

We first project the neuron upon a plane parallel to the long axis of
the cell body (Fig.~(\ref{celllayout}a)).  For a cortical pyramidal
cell, the view would correspond to having flattened out the cortex and
then looking at the cell in plane of the cortical sheet with the axon
and basal dendrites toward the bottom and the apical and distal
dendrites reaching upward. Each point in the figure corresponds to a
current segment in the full, multi-compartmental model of this cell.
It's clear that the cell is not completely symmetric since the left
and right portions, relative to the vertical axis of the cell, do not
exactly correspond.  However, we neglect this anisotropy, and simply
mirror the cell along this axis, averaging any overlapping current
segments (Fig.~(\ref{celllayout}b)).  After performing this simple
transformation, we now assume cylindrical symmetry along the
vertical axis of the cell, with the current elements providing a
current density over the corresponding cylindrical volume.  Viewed out
of plane, the cell would appear as an assortment of cylindrical
annuli.

By symmetrizing the cell, we greatly simplify the problem, since all
terms where $m \ne 0$ integrate to zero in Eq. (\ref{mexp}).  The
remaining $m=0$ spherical harmonics are related to the Laguerre
polynomials, $P_l(x)$, via
$Y_{l,0}(\theta,\varphi)=\sqrt{(2l+1)/(4\pi)}P_l(\cos\theta)$ which
simplifies Eq.~(\ref{mexp}) and the calculation of the moments in
Eqs.~(\ref{moments1},\ref{moments2}).

\section{Inner-Field cellular moments}
We begin our analysis by considering the near-field moments which are
relevant to local intracranial recordings of neural activity .  In the
inner-field, because of the changing volumes of the regions defined by
$R_1$ and $R_2$, both classical and inverse moments are dependent upon
distance.  To efficiently compute the multipole expansion within this
domain, we follow a similar procedure to that outlined in
\cite{rafat}.  The idea is to divide the inner-field into a
series of $N$ spherical shells and to then calculate the classical and
inverse moments in a piecewise fashion within each shell.  This
calculation needs to be performed only once at each time step and may
then be stored within a lookup table.  To calculate the extracellular
field of the cell requires the evaluation of the integrals in
Eqs.(\ref{moments1},\ref{moments2}) which now become sums over the
appropriate subset of $N$ shells, with an interpolation performed at
the boundary between regions $R_2$ and $R_1$. Since the brunt of the
numerics may be performed ahead of time and stored within computer
memory, this method provides an efficient way of calculating the
moments at any radial distance within the inner-field granted that the
expansion converges for a modest number of terms.  For the model
pyramidal cell that we investigate, as an example, we take N=200
shells recorded over 200 time-steps.  To store the first 25 inverse and
classical moments, we must generate a lookup table of approximately 16
Mbytes which can easily be stored in the memory of a modern desktop
computer.

 \begin{figure}[t]
\begin{center}
\includegraphics[width=0.7\columnwidth]{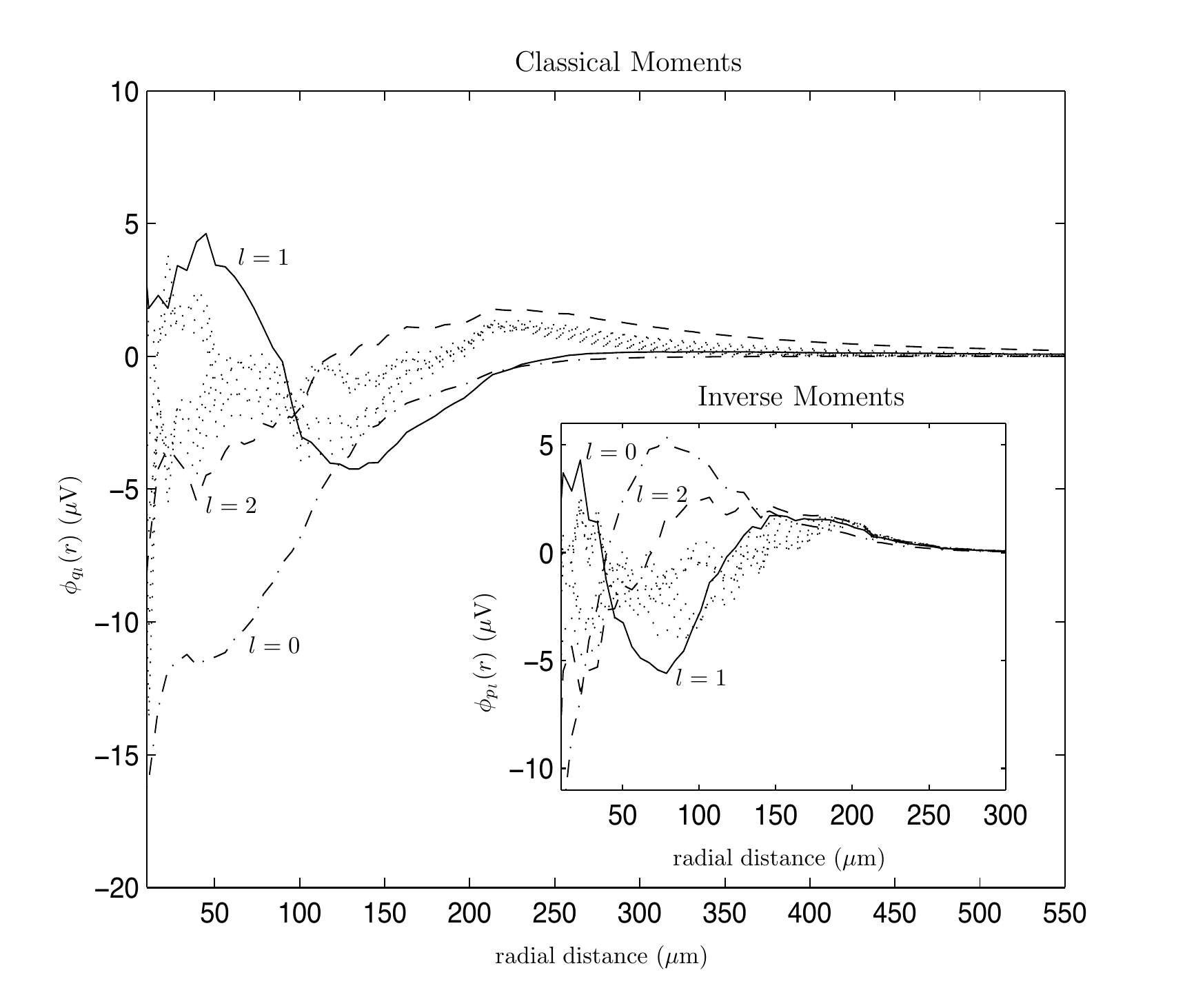}
\caption{The radial potentials $\phi_q(r)$ for the first 11 classical
moments as a function of radial distance at $t=1.4$ ms,
corresponding to the peak of the action potential. The $l=0$ monopole
moment (dash-dotted line), $l=1$ dipole (solid line), and the $l=2$
quadrupole (dashed line) are emphasized along side the remaining
moments up to $l=10$ (dotted lines).  Insert: The radial potential
$\phi_p(r)$ for the first 11 inverse moments (same line labels as
before) .}
\label{nfmoms}
\end{center}
\end{figure}

One might hope that only the first few moments define the
extracellular field of the cell; however, within the near field, the
current distribution is too complex to allow such a simplification and
the moment expansion contains many comparable terms throughout the
time course of the action potential.  We illustrate this in
Fig. (\ref{nfmoms}) for a representative time ($t=1.4\ {\rm ms}$,
corresponding to the peak in the action potential) where, for
clarity, we display only the first 11 classical $\phi_{q_l}(r)$ and 11
inverse $\phi_{p_l}(r)$ radial potentials.

The fairly slow convergence of the weights of the expansion, displayed
in Fig.~(\ref{nfmoms}), is similar for various times about
the action potential. If we exclude a radius of $10-15 {\rm \mu m}$
about the center of the cell, guaranteeing we are
outside the body of the cell itself, the first $
25$ classical and inverse moments are needed to account for the total potential
to within a few percent throughout the entire timecourse of the action
potential.

From Fig.~(\ref{nfmoms}) it is hard to justify any dominant moments of
the cellular current distribution due to the strong radial dependence
displayed. This clearly implies that if we were to model the
extracellular field of this neuron within radial distances on the
order of half the length of the cell ($\sim 550 {\rm \mu m}$), we
must account for the full complexity of the current distribution, and
that any assumption of treating such a complex current distribution
as, perhaps an oscillating dipole, would be unjustified.  Nonetheless,
summing over roughly $50$ elements ($25$ inverse moments and
$25$ classical moments) is a much quicker way to evaluate the
extracellular field than summing over the $\sim 1000$ current sources
of the compartmental model.  We next turn our attention to the
outer-field results.

\begin{figure}[t]
\begin{center}
\includegraphics[scale=0.5]{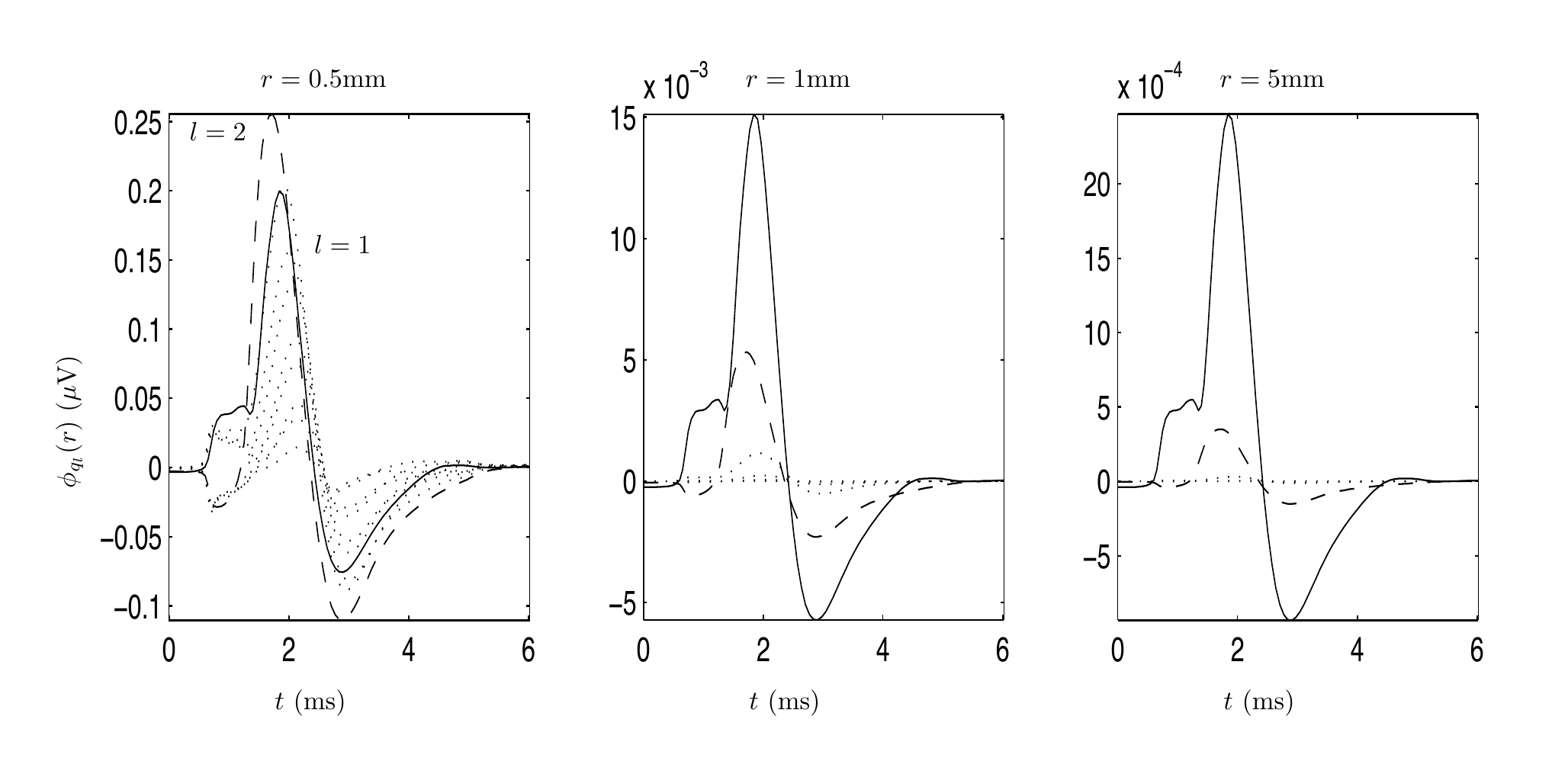} 
\caption{The radial potential $\phi_q(r)$ from the first 11 classical
 moments, as a function of time, at various distances from the cell
 origin. The $l=1$ dipole moment (solid line) and the $l=2$ quadrupole
 moment (dashed line) are emphasized.  For comparison, the higher
 moments (dotted lines), up to $l=10$, are shown.}
\label{momcomps}
\end{center}
\end{figure}

\section{Outer-Field cellular moments}
Within the outer-field, our problem simplifies to a moment
calculation which can be performed without any of the complexities
introduced within the inner-field.  One might assume, since the total
current across the single neuron is conserved (i.e., the $l=0$ moment
is zero), that far from the cell the only significant contributions
would come from the dipole moment ($l=1$).  However, the quadrupole
moment scales only one inverse power of $r$ faster ($1/r^3$ as
compared to $1/r^2$).  If we compare the magnitude of the $1/r^2$
dipole potential to the $1/r^3$ quadrupole potential at a point on the
boundary between the inner and outer-field ($r\sim 0.5$mm, which is
half the length of the cell), in order for the quadrupole component to
remain at say $10\%$ of the magnitude of the dipole component after a
distance of $ \sim 1$cm, ignoring angular dependencies, the initial
magnitude of the quadrupole term at the boundary needs to be only on
the order of twice as large as the dipole term at that same point.
This means that if the magnitude of the quadrupole moment ever exceeds
that of the dipole moment at the boundary to the outer-field, that
there may be a significant region in which the quadrupole moment cannot be neglected.

We stress the above point because our numerical results for the model
pyramidal cell we have been considering displays a rather large
quadrupole moment at various times during the action potential. Figure
(\ref{momcomps}) shows the contribution of the first 11 classical
moments, over the course of the action potential, as we progressively
move away from the cell.  At the cell boundary ($r\sim0.5$mm) the
quadrupole is seen to surpass the dipole moment throughout most of the
timecourse, whereas the higher order moments progressively decay.  At a
radial distance of roughly 1mm, the quadrupole is clearly displayed
while the higher moments are quickly decaying, and at a distance of
$r\sim 0.5$cm only the dipole and quadrupole remain, with a
significant contribution from the quadrupole.   

\begin{figure}[t]
\begin{center}
\includegraphics[width=.75\columnwidth]{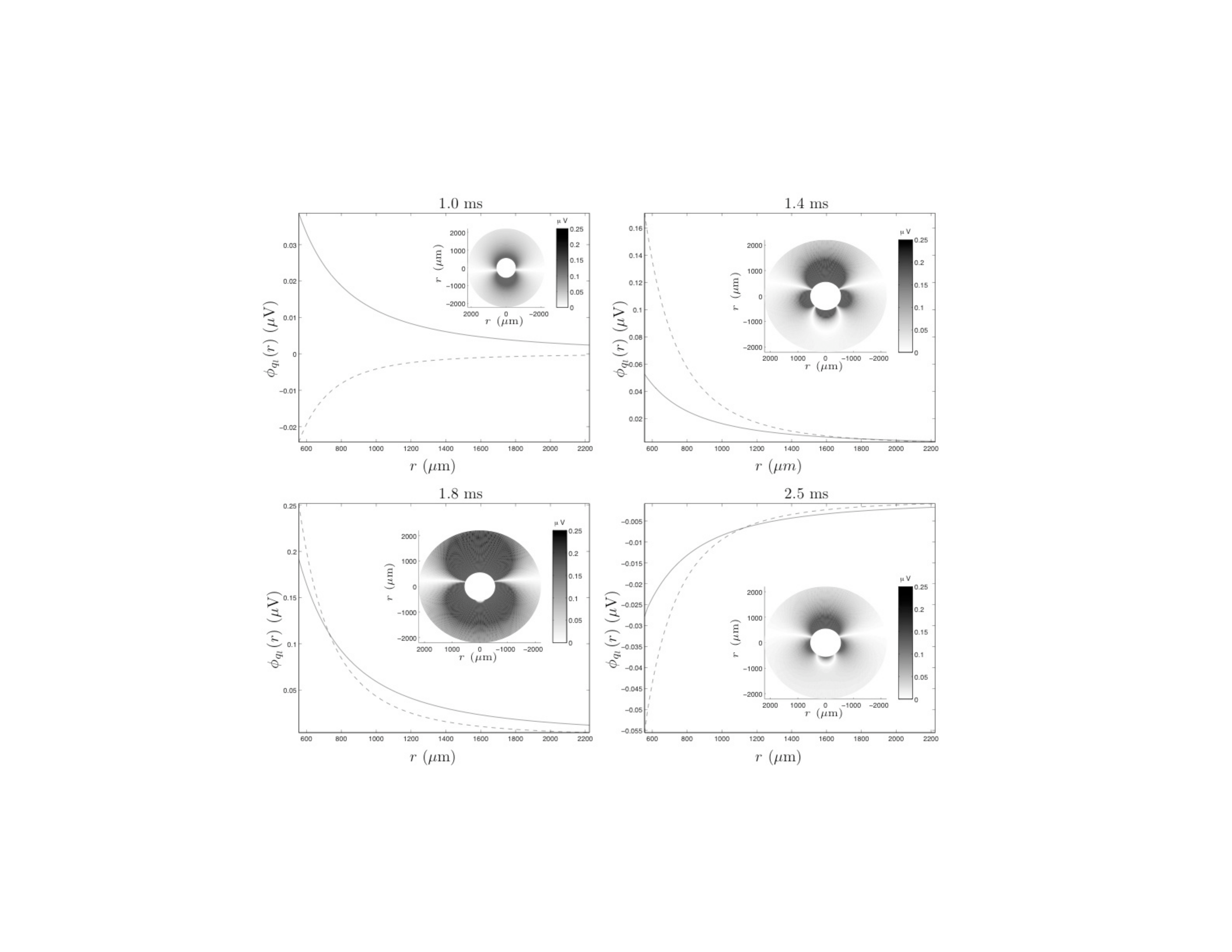}
\caption{Moment comparison in the outer-field, about the peak in the
action potential.  The main figures show only the dominant dipole
$l=1$ (solid line) and quadrupole $l=2$ (dashed line) contributions to
the radial potential $\phi_q(r)$.  The insets show the absolute value of the resulting total
potential in ${\rm \mu V}$. Excluded is the inner-field ($ r < 550
{\rm \mu m}$) where the cell would be oriented along it's vertical
axis as shown in Fig~(\ref{celllayout}). Left to right, starting on
the top, the times are given by $1.0,1.4,1.8$ and $2.5$ms.}
\label{farfieldfigs}
\end{center}
\end{figure}

Figure (\ref{farfieldfigs}) displays the radial contribution to the
extracellular potential from the dipole and quadrupole
moment\footnote{High-resolution color images and animations of the
extracellular potentials can be found at
\url{http://www.klab.caltech.edu/~milstein/moments}}.  As previously
shown, these two moments dominate the extracellular field in the
outer-field throughout the action potential, except for at points
close to the boundary.  We, therefore, neglect the contributions of
all higher moments ($l>2$) in the figures.  At approximately $1\ {\rm
ms}$ into the simulation, the cell begins to spike, and a large dipole
moment dominates.  However, as the action potential grows, a
significant quadrupole moment emerges.  The initial magnitude of this
moment is more than three times that of the dipole which means that it
will contribute to the extracellular field over a significant spatial
extent.  When the action potential has peaked, the dipole moment has
again gained in magnitude, and the extracellular field is clearly
dominated by this moment, which remains until the hyperpolarization of
the cell overshoots the threshold and a relatively strong quadrupole
emerges again, although the overall extracellular potential is much
smaller at this point.

\section{Discussion}

We have shown that the extracellular field of a biologically realistic
pyramidal cell can be accurately and efficiently calculated at all
spatial distances from the cell through a moment expansion
of the membrane current distribution. We have formulated the multipole
expansion in a form that converges at all points in space,
generalizing it from the traditional classical expansion to include
test points localized within the sphere of the current distribution. 

For the model cell under consideration, we have found that we may
divide the extracellular space into three different regions.  In what
we have designated the inner-field, which extends from the
origin--where we have placed the center of the cell--up to length
scales of $0.5$ mm, our analysis has shown that the multipole
expansion converges slowly, requiring on order of the first 25 moments
to converge to within a few percent of the true extracellular
potential.  At slightly larger distances, from just outside the
boundary between the inner and outer field, $r>0.5$mm, the cell
displays a strong quadrupole moment that may appreciably contribute to
the extracellular potential to distances on the order of $1$cm from
the cell.  Within this, region the extracellular field may be modeled
as originating from an oscillating dipole and quadrupole, while higher
moments may be neglected.  At length scales $r > 1 {\rm cm}$, as
expected, only the dipole term remains.

In developing the present method, we have made several assumptions
that should be reconsidered.  First, we have taken full advantage of
the symmetry displayed by pyramidal cells to reduce the number of
terms in the potential expansion--in truth, this is an approximation
as is made clear by Fig.~\ref{cylf}.  It should be noted that for
cells without a clear symmetry axis (e.g. Purkinje cells), one would
have to account for all $m=-l\ldots l$ axial moments.  This would
likely lead to a large number of terms in the potential expansion,
making the present procedure impractical.  Second, we have treated the
extracellular medium as homogeneous, neglecting the effects of other
dendrites or axons present within the vicinity of the cell.  It would
be very difficult if not impossible to exactly account for these
inhomogeneities, nonetheless, it is an interesting question to ask,
for instance, how random defects in the extracellular mileu might
modulate the extracellular field.  Third, we have triggered the action
potential within the soma and have analysed the extracellular field
generated by the dynamics of the resulting membrane currents.  One may
also initiate the action potential by distributing the imputs within
the synapses and proceed with the analysis we have presented here.
Since the present method will work for an arbitrary current
distribution, only the efficiency of our method should be effected.

As discussed in the introduction, the contribution of the action
potential is thought to be negligible to EEG measurements.  We are now
in a better position to test this fundamental assumption.  For
instance, we may use the method presented here to simulate large
populations of biologically realistic spiking neurons and see the
effects of orientation and synchrony on the combined extracellular
potentials.  In particular, we may study the contributions of slower
components following the action potential, such as short and
longer-lasting after-hyperpolarizations.  Unfortunately, due to the
complexity of the current dynamics displayed by the model neuron we
have used for this study, it is difficult to infer how these slower
processes would effect the extracellular fields without fully
simulating these fields.

\mbox{}\\ \\ We would like to thank Carl Gold for providing us with his
NEURON code package.  Joshua Milstein and Christof Koch acknowledge
support from the Swartz Foundation and NSF.

\end{document}